\documentstyle[11pt,epsfig]{article}

\newcommand{\et}{\sl et al \rm}

\newcommand{\um}{\mbox{$\mu\rm m$}} 
\newcommand{\uJy}{\mbox{$\mu\rm Jy$}}

\newcommand{\arcsec}{\mbox{$^{\prime\prime}$}}

\newcommand{\vol}[2]{\bf #1, \rm #2.}

\newcommand{\apj}{\sl Astrophys.J. \rm}
\newcommand{\apjl}{\sl Astrophys.J.Lett \rm}  
\newcommand{\mn}{\sl Mon.Not.R.astr.Soc., \rm} 
\newcommand{\AnA}{\sl Astron.Astrophys., \rm}

\newcommand{\aj}{\sl Astron.J. \rm} 
\newcommand{\anrev}{\sl Ann.Rev.Astron.Astrophys., \rm}
\newcommand{\nat}{\sl Nature, \rm}

\begin{document}


\begin{center} 
\vskip 15pt
\LARGE Blank field submm sources, failed stars, and the dark matter \\
\Large 
\vskip 1cm

Monthly Notices of RAS, in press. \\
\large
(original submitted March 20th 2000) \\
\Large
\vskip 1cm

A.Lawrence \\ 
Institute for Astronomy, University of Edinburgh \\
Royal Observatory, Blackford Hill, Edinburgh EH9 3HJ  \\ 
\vskip 0.3cm

\normalsize \end{center}


\baselineskip 12pt 

\vskip 30pt
{\bf ABSTRACT}

I discuss the possibility that a significant fraction of the extremely
common faint submm sources found in recent surveys are not in fact high
redshift galaxies, but actually local objects emitting only in the submm, with
a temperature around $7K$. The majority of faint SCUBA sources clearly
really are distant galaxies.
However if even a quarter or a third of the SCUBA sources are actually local
objects, the cosmological implications are significant, as this would
selectively remove the objects believed to be at $z>3$.
Two hypotheses - very cold brown dwarfs, and outer
solar system bodies - are easily rejected. A third hypothesis - cold dark
dusty gas clouds - is not so easily dismissed. I show that the observational
constraints on such a population - dynamical limits on local missing matter,
the FIR-mm background, and the absence of gross high-latitude extinction
features - constrains the mass of such objects to be in the mass range $0.1$ to
$10$ Jupiter masses. On the assumption of virial equilibrium, their sizes are in
the range $1 - 100$ AU, with angular sizes around a tenth of an arcsecond. They
would be completely opaque at visible and IR wavelengths. The characteristics
deduced are closely similiar to those of the objects proposed by Walker and
Wardle (1998) to explain ``extreme scattering events" in quasar radio light
curves, and which they propose fill the Galactic halo and explain halo dark
matter. Indeed, at around $1$ Jupiter mass, the local population density would
be similar to that in dark halo models. However, such objects, if they explain
a large fraction of the SCUBA submm sources, cannot extend through the halo
without greatly exceeding the FIR-mm background. Instead, I deduce the
characteristic distance of the SCUBA sources to be around $100$ pc, consistent
with being drawn from a disk population with a scale height of a few hundred
parsecs. Possibly a ``Population II" dust-less version of such objects could
exist in the halo. Regardless of the dark matter problem, the possible existence
of such compact sub-stellar but non-degenerate objects is intriguing. Such
objects should collapse on a very short timescale, but at such a low
temperature it is possible that cosmic ray heating can maintain them in
equilibrium. The main theoretical objection is that such an equilibrium may be
unstable on a thermal timescale. If however such objects do exist, they may be
seen as  ``failed stars", representing an alternative
 end-point to stars and brown-dwarfs. It is possible that they greatly
outnumber both stars and brown dwarfs. The nearest such object could be a
fraction of a parsec away. Several relatively simple observations could
critically test this hypothesis.



\section{Introduction}  \label{intro}

Deep submm surveys of small patches of sky have only just become feasible,
largely due to the development of the SCUBA instrument on the James Clerk
Maxwell Telescope (Holland \et\ 1999). Several groups have pursued surveys on
scales of a few arcminutes to mJy levels (Smail, Ivison, and Blain 1997; Hughes
\et\ 1998; Barger \et\ 1998, 1999; Eales \et\ 1998; Lilly \et\ 1999; Blain \et\
1999a), finding a very large sky density of faint submm sources. These surveys
seem to have justified the long held belief that submm surveys would be a
short-cut to the high-redshift universe, as the so-called ``negative
k-correction" effect largely counterbalances the dimming due to distance and
redshift, so that objects can be seen just as easily at very high redshift
($z=3-10$) as they can at moderate redshift ($z\sim 1$). High redshift optical
identifications have indeed been proposed for many of the submm sources in the
above papers. The IR-mm luminosities of the sources are then very large, similar
to local ultraluminous IRAS galaxies, with implied star formation rates of
hundreds of solar masses per year. However, given the sky density of these
sources, the high redshift luminosity density must be large. Hughes \et\ (1998),
in their study of submm sources in the Hubble Deep Field (HDF), show that the
implied star formation rate at $z=2-5$ is {\em larger} than that deduced from
the UV fluxes of all the optically selected high-z galaxies in the HDF. The
optically derived star-formation history curve, or Lilly-Madau plot, has since
undergone revision but the qualitative point remains unchanged. This has
striking consequences - most of the star formation in the young universe was
obscured, and most of the star formation in the young universe went on in a tiny
minority of starburst galaxies, not in the great majority of quiescent or
moderately active objects. This is quite the reverse of the situation today.
There may be problems with such a picture - Blain \et\ (1999a) show that a
star-formation history of the kind deduced by Hughes \et\ may, over the history
of the universe, overproduce the metals, and overproduce the number of observed
stars.  

It may very well be that the above story is broadly correct, but it is so
important that we must make sure that we have excluded reasonable alternatives.
One such alternative is that some fraction of the submm sources are actually AGN
rather than starbursts (Almaini Lawrence and Boyle 1999;  McMahon \et\ 1999;
Fabian \et\ 2000;  Gunn \et\ in preparation). Another more radical alternative is
that the submm surveys, like all other first surveys of the sky at a new
wavelength, are discovering a {\em new class of object}. We are driven to
examine such an idea because the optical identifications of submm sources with
high redshift galaxies, while highly plausible, are often only circumstantial.
The problem of course is the large beam size of SCUBA (14\arcsec ) so that
identifications are nearly always ambiguous. In some cases, a relatively bright
galaxy is within the error circle, and so statistically secure. But these are
the less interesting cases - reasonably secure identifications at $z\sim0.5-1$. 
In some other cases, identifications in the range $z\sim 1-3 $ are secure
because of confirming CO detections, or because of the presence of extremely
red objects (EROs) which are relatively rare (Ivison \et\ 1998; Frayer \et\
1998, 1999; Smail \et\ 1999). In a large number of cases however, several faint
blobs are positionally consistent, and the identification rests on the
constraints given by limits on the spectral energy distribution (SED) - for
example non-detection at 15\um\ in the ISO data together with a detection at
850\um\ ruling out objects with an ARP~220-like SED at low redshift (see Hughes
\et\ 1998), or the submm/radio ratio (Carilli and Yun 1999; Blain 1999). In all
such cases, the submm source is equally consistent with having no optical
counterpart at all.

For the brightest SCUBA sources, mm-wave interferometry is just feasible and
can of course substantially increase locational accuracy. Downes \et\ (1999)
observed HDF~850.1, the brightest SCUBA source in the HDF, with IRAM, and
achieved a position accurate to 0.3\arcsec . Within the $2\sigma$
contour, there is {\em no object at all} to the limit of the HDF. Two very faint
objects are consistent at $3\sigma$. However, the tentative photometric redshifts
of those objects are inconsistent with the SED of HDF~850.1, if the intrinsic
SED is like that of ARP~220 or M82 (Hughes \et\ 1998). There are several
possible explanations. (i) It could be that the
intrinsic SED is different from those classic exemplars, and indeed Downes \et\
find a local object whose SED does appear to be consistent. (ii) It could simply
be that the tentative redshifts are wrong. (iii) The most interesting possibility
is that the correct identification is at very high redshift, $z>5$. The
combination of distance dimming, k-correction, and some reddening could move
such an object out of the detection limit for the HDF. Deep K-band imaging by
Smail \et\ (1999) has shown that perhaps 10\% of the SCUBA sources are
``Extremely Red Objects (EROs)", appearing only in the IR. Unpublished deep
K-band imaging of the CFRS survey sources show that only one third appear to
identified, even to $K=21.5$, suggesting that {\em most} counterparts are at
$z>2$ (Eales, private communication).

In some ways the simplest explanation for the non-identification of HDF~850.1
is that it is an object that is bright in the submm but emits negligible light
at optical wavelengths. Is it possible that such a previously
unsuspected population of objects exists, and at such a high sky density  ? We
must at least discount such a hypothesis before believing the consequences of
submm surveys for cosmology. In this paper I consider three such possible
populations and test them against the constraints we have. The first two -
brown dwarfs and solar system objects - are quickly dispensed
with. The third hypothetical population examined - very cold dense dark dust
clouds - turns out to be much harder to dismiss, as long as the clouds are in
the sub-stellar mass range. Furthermore such objects may plausibly contribute
substantially to the dark matter problem, as has been suggested by other
authors for quite different reasons (Pfenniger Combes and Martinet 1994; Gerhard
and Silk 1996; Walker and Wardle 1998). Once we have arrived at the concept of
such objects, they seem more than plausible, as a likely end-product of cloud
collapse, and as an alternative to brown dwarfs. Even if they do not turn out to
constitute a large fraction of the SCUBA sources, it is of some interest to
discover whether they exist at all. 

\section{Observational constraints}  \label{constr} 

Any model aiming to explain the sub-mm sources must satisfy a substantial
number of constraints. (i) The objects must have an SED consistent with what
we know from the brighter submm sources. (ii) Their distribution on the
sky must be at least grossly isotropic, although current surveys cannot
exclude variations of the order of tens of percent. (iii) The deduced
space density must not exceed the limits imposed by local dynamics. (iv) Their
integrated surface brightness must not exceed the FIR-mm background discovered
by COBE. (v) The population covering factor must be small, or the objects
would already have been discovered through extinction effects. Most of these
constraints will be revisited as we proceed, but I now look at two of the
key constraints more carefully, followed by a brief discussion of the possible
importance of radio emission.

\subsection{Temperature constraint from the SED of HDF~850.1} \label{temp}

To constrain the SED, I have taken HDF~850.1 as the exemplar, as it has no
identification, but has been measured at a number of wavelengths. Table 1 lists
the reported fluxes and limits for HDF~850.1. Table 2 takes various ratios and
for each lists the range of temperatures consistent with the ratio concerned, for
(a) a blackbody
 SED, (b) a greybody SED with various values of $\beta$. Here the greybody form
is defined as usual as a spectrum with the form $F_\nu=Q_\nu B_\nu(T)$ where
$Q_\nu \propto \nu^\beta$. The limits and ranges discussed below are all
effectively at approximately 95\% confidence. 

The crucial ratio is S(450\um )/S(850\um ). This has an effective power-law
slope $\alpha<1.2$, much flatter than a Rayleigh-Jeans slope, telling us that
whether blackbody or greybody, the objects must be extremely cold - blackbodies
must be colder than 18K, and greybodies colder than 9K. The long wavelength
ratios are also interesting. The ratio S(1300\um )/S(850\um ) has an effective
power law slope of $\alpha=2.7\pm0.4$, favouring a greybody interpretation but
not quite ruling out a blackbody. The ratio of S(2800\um )/S(1300\um ) implies
a power law slope of $\alpha>2.3$, formally ruling out blackbody emission. Given
the likely systematic errors on these fluxes, we cannot be confident of
this conclusion, but a greybody is clearly preferred.

Overall, blackbody emission is mildly ruled out, and must in any case be
colder than 18K. For greybody emission, we cannot separate variations in
$\beta$ and $T$, but for a given $\beta$ the implied $T$ is very tightly
constrained. For $\beta$ anywhere within the normally considered range for
Galactic dust ($\beta\sim 1-2$), the total allowed range of temperature is
$T=4.7 - 9.4$. For much of this paper, I take $T=7$ as the best-bet temperature
value. 

Most other blank field survey SCUBA sources have so far been measured only at a
single wavelength, 850\um . They are generally too faint to detect at 450\um .
However Eales \et\ (2000) have co-added the 450\um\ data at the
position of their 850\um\ sources, and find that the mean value of the flux
ratio at these two wavelengths is S(450)/S(850)$<$ 1.9, at 3$\sigma$
confidence. For a variety of reasons discussed in their paper, there may
systematic uncertainties in this analysis, so that the true ratio could be
as large as 3. However it is clear that HDF~850.1 is at least not unnusual, 
and it is even possible that most sources are even colder.

\subsection{Limits on space density of any unknown population}

It is possible that the kinds of objects we will consider are indeed submm
sources and of astrophysical interest, but we are primarily concerned with
whether such objects are {\em common} enough to explain a substantial
fraction of the sources in the blank field surveys. The number counts from
submm surveys still have considerable uncertainty, but down to S(850\um ) = 2
mJy the surface density of sources is $N \sim 2000-3000$ deg$^{-2}$, i.e. a
source every arcminute or two (Hughes \et\ 1998; Blain \et\ 1999a; Barger, Cowie
and Sanders 1999). If any new population is to be significant, then it must have
let us say $N \sim 1000$ deg$^{-2}$ at a flux of 2 mJy. This surface density
will be taken as a reference point in everything that follows. If we postulate a
population of objects with a given luminosity, we can calculate the distance $D$
to which they are detectable to 2 mJy. If we further assume a mass $M$ for the
objects, then we can deduce the overall mass density of the population,
$\rho=3MN/D^3$. 

In the solar neighbourhood we now have a reasonably agreed census for both
stars and interstellar matter, and in addition have an estimate of the
dynamical mass from vertical motions of stars (see eg Binney and Merrifield
1999 and references therein; and the most recent determinations from Creze \et\
(1998) and Holmberg and Flynn (2000). The local dynamical mass and the
observed
material are in quite good agreement at around $\rho \sim 0.1 M_\odot$
pc$^{-3}$. Of this, roughly 45\% is in stars and remnants, and 55\% in
gas. The
uncertainties are still significant, but a fairly robust conclusion is that any
hitherto undetected component in the solar neighbourhood cannot have a density
greater than $\rho\sim 0.03M_\odot$ pc$^{-3}$. Note however that this limit is
still quite consistent with the likely local contribution from halo dark matter;
a variety of models is still consistent with the data, but a solar neighbourhood
normalisation of around $\rho \sim 0.01 M_\odot$ pc$^{-3}$ is quite typical (eg
Alcock \et 1996; Evans 1996).

\subsection{Possible radio emission}

Although the IRAM position of HDF~850.1 coincides with no optical counterpart,
there is a marginally significant and {\em extremely faint} radio source
which, if indeed real, is nicely consistent with the position of the submm
source. (See Figure 4 of Downes \et ). The ratio S(850\um )/S(3.5cm)
$\sim$ 900,
some two orders of magnitude larger than that seen in local starburst galaxies,
but quite possibly consistent with that expected from a highly redshifted
starburst (Carilli and Yun 1999; Blain 1999). At lower frequency the source
is not detected, so that S(850\um )/S(20cm) $>$ 300. Smail \et\ (2000)  found
faint radio counterparts to 7/15 SCUBA sources that they surveyed, with nearly
all of these being the brightest submm sources and/or those with reasonably
secure optical counterparts. The mean flux ratio for the detections was
 S(850\um )/S(20cm) $\sim$ 140, consistent with tentative redshifts in the
range z=1-3. Including the undetected sources, Smail \et\ state that a
conservative value for the median redshift is $z\geq 2.5$. Another deep radio
study, of the CFRS survey SCUBA sources, detected 5/19 objects (Eales
\et\ 2000). The majority of the faintest SCUBA sources are therefore
either
at very high redshift (z$>$3) or are simply not radio sources, or at least
extremely feeble ones.

In summary, it may be that there is no effect to explain, but potentially the
existence of, albeit extremely weak, radio emission is a difficult challenge
to local hypotheses for the SCUBA sources. We return to this question in
section \ref{radio}.

\section{Brown Dwarfs}  \label{subs} 

Brown dwarf candidates discovered in the last few years are bright IR sources
but optically faint, with temperatures of the order 1000K. Is it possible
there is a large population of low mass and/or very old sub-stellar degenerate
objects that are completely invisible at optical wavelengths ? Presumably
their gross SED would be at least approximately blackbody, which may be ruled
out in the case of HDF~850.1. More importantly, current brown dwarf models
(Burrows \et\ 1997) find that even for Jupiter sized objects that are as old as
the Galaxy, the expected temperature is still of the order 90K, quite
inconsistent with the observed SED of HDF~850.1. Even ignoring this problem, we
can soon see that the population can only be seen very close by, and would have
an absurdly large space density. Scaling to the radius of Jupiter ($R_J$), we
can estimate the distance at which such an object would have an 850\um\ flux of
2 mJy :

$$  D({\rm pc}) = 0.018\times \left(\frac{R}{R_J}\right)
\left(\frac{T}{10}\right)^{1/2} \left(\frac{F}{2{\rm mJy}}\right)^{-1/2}   $$

I have assumed blackbody radiation at $T\sim 10K$, and taken the Rayleigh-Jeans
approximation. If we now assume a mass $M$ for each object, also scaled to the
Jupiter value $M_J$, then for a sky density of $N=1000$ deg$^{-2}$ we get an
overall population space density

$$ \rho(M_\odot {\rm pc}^{-3})  = 1.63 \times 10^9 \times
\left(\frac{M}{M_J}\right) \left(\frac{R}{R_J}\right)^{-3}
\left(\frac{T}{10}\right)^{-3/2}  \left(\frac{F}{2{\rm mJy}}\right)^{3/2} 
\left(\frac{N}{1000}\right) $$ 

We conclude that a population of brown dwarfs cannot explain the majority of the
SCUBA sources. If such very old small brown dwarfs do exist at a temperature of
$T\sim 100K$, then at a flux level of 2 mJy there can only be a handful over the
whole sky without violating the local mass density limit.

Examining the above equation, we can see that the escape route is that for a
given mass one wants a much larger radius. To be consistent with local density
limits, Jupiter mass objects would need to be a few AU in size.  We examine
this possibility shortly. But first, another possibility is to find a locality
where the density might just be that high - inside the solar system.

\section{Trans-Neptunian Objects}  \label{TNOs} 

In recent years wide field searches have discovered that there is a
substantial population of asteroid-sized bodies beyond Neptune (eg
review by Weissman 1995). These are often referred to as ``Kuiper Belt Objects"
although their latitude and orbit distribution is not yet well known. Beyond
this, it has of course been long postulated that there is a vast population of
icy bodies which acts as the reservoir for comets (i.e. the ``Oort Cloud"),
perhaps as many as $10^{13}$ objects out to 30,000 AU. This is of course more
than enough sky density but we are unlikely to detect objects that far. If we
assume an albedo of 0.1 (typical of comets) then at a distance $D$ solar
heating leads to a temperature

$$ T_{ss} = 38.5 \left(\frac{D}{100 {\rm AU}}\right)^{-1/2} $$

Here I have used the ``sub-solar temperature" appropriate to the surface facing
the sun, as this is also the face that radiates back toward us. To be as cold
as $T\sim 7K$ the object would need to be at a distance of $\sim 3000$ AU. To
have a submm flux of 2 mJy an object would then need a size of around 70,000
km, i.e. such objects would be of roughly Jupiter size. If the sky density is as
large as $N=1000$ deg$^{-2}$, then the total mass in such objects would be of
the order of $10^4 M_\odot$ - unlikely not to have been noticed. If we put aside
the temperature constraint and just calculate the submm flux, then we find that
an object of radius $R$ would be detectable to a distance given by

$$  \left(\frac{D}{100{\rm AU}}\right) =  1.01 \times \left[\left(\frac{R}{1000
{\rm km}}\right)^2/\left(\frac{F}{2 {\rm mJy}}\right)\right]^{2/5}  $$

Thus a Pluto sized object ($\sim$ 1000 km) could be seen to $\sim$ 100 AU,
whereas an asteroid sized object ($\sim$ 100 km) could be seen to $\sim$ 16
AU. This roughly matches the size and distance range of
the trans-Neptunian objects seen so far, to limiting magnitudes
around $V\sim 25$. Such known TNOs should then be just visible to SCUBA. However,
the surface density of objects to V$\sim$ 25 is $\sim$10 deg$^{-2}$, so they fail
to explain a significant fraction of SCUBA sources by around 2 orders of
magnitude. The final blow is that at such distances objects still have have
significant parallactic motion. (This is after all how they have been
discovered). Even at 100 AU the expected motion is 32\arcsec\ per day. However
the 2mJy depth achieved by Hughes \et\ (1998) in the HDF required 50 hours of
integration spread over many separate observing nights. 

I conclude that outer solar system bodies cannot explain a significant
fraction of discrete SCUBA sources. It remains possible of course that the
integrated emission contributes to the FIR background. However Dwek \et\ (1998)
go through similar calculations and conclude that solar system objects are not a
significant contributor to the background.

\section{Cold Dark Clouds}  \label{CDCs} 

We know of course that cold dark clouds exist in the Galactic plane, and they
emit strongly in the submm.  The question here is whether they can be small
enough, cold enough, and common enough (at high as well as low galactic
latitudes) to explain the population of discrete SCUBA sources. I consider
clouds of mass $M$, with gas-to-dust ratio $\mu$, emitting in the submm by
optically thin grey-body emission from dust particles at temperature $T$. (In a
later section I check the optically thin assumption, find that it may be invalid
for some of the mass range considered, and repeat some of the key calculations
assuming blackbody emission). The mass of dust can be computed from the observed
flux in the usual way from

$$ M_d = \frac{F_\nu D^2}{K_d(\nu) B_\nu(T)} $$

Here $D$ is the distance of the cloud, $B\nu(T)$ is the Planck function, and
$K_d(\nu)$ is the usual dust emission coefficient, $K_d(\nu)=3Q_\nu/4a\rho$,
where $Q_\nu$ is the dust emissivity, $a$ is the grain radius, and $\rho$ the
grain density. For a wavelength of 850\um , we use $K_d=0.14$ (see for example
Hughes, Dunlop, and Rawlings 1997). 

\subsection{Distance and space density} \label{distance}

We can now calculate the distance at which a cloud of given mass will have a
flux of 2 mJy. I quote the result scaled to a Jupiter
mass, as we shall see shortly that this is the relevant mass scale. The distance
derived is proportional to the square root of the Planck function. A little
numerical experimentation shows that at a wavelength of 850\um\ an
approximation accurate to 10\% over the range $2-12K$ is  $\sqrt{B\nu(T)}
\propto (T-2)/5 $. I use this approximation to show the effect of
varying the assumed temperature, normalised to a value of $T=7K$. The result is 

$$ D({\rm pc}) = 94 \times \left(\frac{M}{M_J}\right)^{1/2}
\left(\frac{\mu}{100}\right)^{-1/2} \left(\frac{T-2}{5}\right)
\left(\frac{F}{2 {\rm mJy}}\right)^{-1/2} $$

Given this result, there are two appealing mass scales. For $M\sim10^4 M_J\sim
10M_\odot$ the clouds would be at 10kpc, and could thus be a halo population and
so isotropic. At $M\sim M_J$ the characteristic distance is of the order
100 pc. Such objects could then a disk population, but are close enough to
be approximately but not precisely isotropic. (Of course fainter objects could
be further away, but in section \ref{FIRB} we show that in fact the population
cannot extend much further). If we now assume that
the population to this distance has surface number count $N=1000$ deg$^{-2}$ we
can calculate the implied mass density

$$ \rho(M_\odot {\rm pc}^{-3}) = 0.011 \times 
\left(\frac{M}{M_J}\right)^{-1/2}
\left(\frac{\mu}{100}\right)^{3/2}
\left(\frac{F}{2 {\rm mJy}}\right)^{3/2}
\left(\frac{T-2}{5}\right)^{-3}
\left(\frac{N}{1000}\right)                         $$ 

If the implied density is to not to exceed the limit on previously undetected
material ($\rho = 0.03$) then we find that $M>0.1 M_J$. Thus extremely small
objects are excluded. If $M\sim M_J$ the implied density is approximately
equal to the expected local density of the dark halo, leading to the
intriguing possibility that SCUBA sources are the famous dark matter. If
$M>>M_J$ the objects concerned are still of great interest, but not
dynamically significant, and not a major component of the interstellar medium
in mass terms.

The number density of our objects, $n=\rho/M$ is found to be :

$$ n({\rm pc}^{-3}) = 11.9 \times 
\left(\frac{M}{M_J}\right)^{-3/2}
\left(\frac{\mu}{100}\right)^{3/2}
\left(\frac{F}{2 {\rm mJy}}\right)^{3/2}
\left(\frac{T-2}{5}\right)^{-3}
\left(\frac{N}{1000}\right)       $$

If the objects are massive (e.g. $M\sim 10M_\odot\sim 10^4M_J$ ) then they are
fairly sparse. However if the objects are small ($M\sim M_J$) then they are
{\em an order of magnitude more common than stars}. While the typical 2 mJy
SCUBA source may be 100 pc away, the nearest object may be only half a parsec
away and as bright as 10 Jy.

\subsection{The FIR-mm background and the spatial distribution
of the clouds} \label{FIRB}

Taking our fiducial values of $N=1000$ deg$^{-2}$ at $F=2$ mJy, and assuming a
simple Euclidean counts model, the summed surface brightness of all sources down
to this flux level should be $I=0.02$ MJy sr$^{-1}$. This is
approximately one seventh of the isotropic background measured by COBE (Fixsen
\et 1998). This is not a surprise, as Hughes \et (1998) have already pointed
out that the SCUBA sources as whole down to 2 mJy explain about one half of the
observed background, and we are working on the hypothesis that the postulated
local population is a half or somewhat less of the total SCUBA source
population. The crucial implication is that the objects concerned cannot stretch
much further in distance than we have already deduced for 2 mJy sources. For
high-redshift galaxies, this is really an effect in time, as more distant
objects are seen at an early epoch. For our postulated local objects, if we
assume that they should contribute no more than say half the background, then,
again assuming a Euclidean counts model, the population can extend no further
than about 3 times the distance deduced for 2 mJy sources. Whether this is
reasonable depends on the mass of the objects. 

(i) If our 2 mJy  cold dark clouds have $M\sim 0.1 - 1 M_J$ they are at a
distance of $\sim 30 -100$ pc, and so cannot extend much beyond $100 - 300$ pc.
However the local stellar disc has a scale height of 300 pc, and the
interstellar medium of around 100 pc (Binney and Merrifield 1999). Objects in
this mass range are therefore nicely consistent with being a disk population.

(ii) If our 2 mJy  cold dark clouds have $M\sim 1-10 M_\odot \sim 10^{3-4} M_J$
then they are at a distance of $\sim 3 - 10$ kpc, and cannot extend beyond $10 -
30$ kpc. This is at least roughly consistent with being a halo population.

These two mass ranges then seem to allow spatial distribution models that are
at least reasonable, as either disk or halo objects. We cannot strictly rule
out objects with $10-100 M_J$ or $M>10^5 M_J$, but the spatial distributions
required are rather arbitrary. 

As well as being consistent with the observed intensity of the FIR-mm
background, our cold dark clouds should also be consistent with its spectrum.
Fixsen \et (1998) found that the spectrum they derived could be well fitted
with a grey body functional form, with $\beta=0.64$ and $T=18.5$. They
intended this as a parametric description of the data rather than a physical
model, and of course an emissivity index $\beta=0.64$ is considerably flatter
than the $\beta \sim 1 - 2 $ normally found in both Galactic and extragalactic
objects. Dwek \et (1998) model the combined FIRAS-DIRBE data by the FIR
emission from star formation over the history of the universe. A redshifted
greybody is still a greybody, with $\beta$ unchanged but the apparent
temperature reduced by a factor $(1+z)$. The flat observed $\beta$ results from
the summation of contributions from a wide range of redshifts. 

Figure 1 shows how the Fixsen \et spectrum can be decomposed into warm and
cold components. I fixed the emissivity index for both components
to be $\beta=1.3$, which is the mean value found in a recent submm study
of 104 IRAS galaxies (Dunne \et\ 2000), and is also fairly typical of
starburst galaxies (e.g. M82 : Hughes \et\ 1994).
The temperature of the cold component was
fixed at $T_2 = 7K$. There were then two free parameters - the temperature $T_1$
of the warm component, and the relative strength of the two components.
There  are
several important features that come out of this analysis. (i) The FIR-mm
background can easily accomodate such a very cold component in a natural
way. The cold component is similar to that claimed to be
seen throughout the Galaxy by Reach \et (1995).
(ii) The warm component has a temperature of around $T_1\sim 17$. This is quite
consistent with an M82-like starburst spectrum as found by Hughes \et , with
$T\sim 50$ and $\beta=1.3$ at $z=2$. (iii) The two components cross at about 800
\um . Note that this was not fixed in advance by the modelling, but seems to be
required to fit the spectrum. Thus at far-IR wavelengths the warm objects
(presumed high-z galaxies) dominate; at mm wavelengths the cold objects
(postulated here to be cold dark clouds) dominate; but at submm wavelengths they
are roughly equally numerous. 

Obviously much more careful modelling is needed, using for example a
template starburst SED rather than a single temperature greybody, and
summing over an
assumed star formation history, as Dwek \et (1988) did. Compared to the Dwek
\et analysis the effect of adding a very cold component is likely to be (a)
requiring a more peaked star formation history, and/or (b) implying a narrower
SED, one more like M82 than the Dwek \et template, ARP~220. Finally however we
note that both the Fixsen \et (1998) and the Puget \et (1997) derivations of
the FIR-mm background required the modelled removal of a Galactic component
assumed to be of fixed colour. This should now be seen as an unsafe
assumption, so the extragalactic background should really be re-derived
self-consistently.

\subsection{Physical size of clouds} \label{phys}

So far I have derived constraints on the allowable mass range based
on local
density limits and the FIR-mm background. The other strong constraint we want
to examine is the covering factor of such objects, for which we need some
estimate of their size. This also allows us to consider other interesting
physical characteristics such as density and column density, and  allows
us to check the optically thin assumption. 

The sources concerned have been found as point sources in SCUBA maps, and so
cannot be very much larger than the SCUBA beam size, 14\arcsec. As they are
rather weak sources, it is probably hard to exclude them having diameters of up
to say 30\arcsec\ across. At the distance derived above for 2 mJy sources, this
corresponds to a physical radius of $R({\rm m}) \leq 2.11\times 10^{14}
(M/M_J)^{1/2}$. For our exemplar HDF~850.1, we can place a tighter limit.
Downes \et\ examine the beam profile and state that the source diameter is less
than 2\arcsec . This is consistent with the fact that the
1.3mm flux seen within the IRAM beam of $2.1\times 1.7\arcsec$ is consistent
with the flux measured at 1350\um\ by Hughes \et (1998) with SCUBA, showing
that the source is not significantly larger than the IRAM beam. We take this as
implying that the angular radius is less than 1\arcsec .  The 850\um\ flux of
HDF~850.1 is 7.0 mJy so that its implied distance is  50.2 pc. At this
distance the physical size of the source is $R{(\rm m)} \leq 7.51\times 10^{12}
(M/M_J)^{1/2}$. 

How big might we {\em expect} such clouds to be ? One possible simplifying
assumption is that they are in virial equilibrium, which is also the same size
(approximately) at which they might be in hydrostatic equilibrium, and at one
Jeans mass. The potential energy will be $U = \alpha G M^2/R$ where $\alpha$
depends on density profile. For uniform density, $\alpha = 3/5$. However we
expect that the clouds are likely to be isothermal (see section \ref{f-stars})
so that $\rho \propto 1/r^2$ which gives $\alpha =1$. Next we assume that the
cloud is almost entirely molecular hydrogen. At a temperature of $7K$ the
rotational levels of $H_2$ will not be excited, so the energy per molecule is
$3kT/2$. Finally we arrive at a virial size

$$ R_{virial}({\rm m}) = 1.46 \times 10^{12} \times
\left(\frac{M}{M_J}\right)
\left(\frac{T}{7}\right)^{-1}    $$

For large masses ($>10^{2-4} M_J$) the observed limits imply a size comparable
to the virial size or even smaller, depending on whether one takes the general
SCUBA limit or the IRAM limit appropriate to HDF~850.1. For smaller clouds the
observed limits allow the clouds to be larger than the virial size. However a
cloud above its virial size will collapse on a free-fall timescale, giving a
collapse time of $t_{collapse} \leq 2,000\, {\rm years} \times (M/M_J)^{1/4}$.
Such objects would soon reach their virial sizes. However we then
arrive at a second problem. The thermal energy content of the cloud is roughly
$M/2m_p \times 3/2 kT$. Its luminosity we can crudely estimate as $\nu L_\nu$
where $L\nu$ is the monochromatic luminosity for a flux of 2mJy and a distance
of 94 pc, with all the usual scalings. We then find that the cooling time is
$t_{cool} \sim 3500\, {\rm years}$, independent of mass.  Maintaining such
clouds in equilibrium therefore requires a heating source, and it is not
obvious whether such an equilibrium would be stable. We return to these
problems in section 8, assuming for the while that they can be solved.

Our postulated clouds are extremely dense compared to normal molecular clouds,
but still very diffuse compared to stars. A Jupiter mass cloud has a size of
9.8 AU, comparable to the orbital radius of Jupiter. The density, in hydrogen
atoms per unit volume, is

$$ n_{H}({\rm m}^{-3}) = 8.72 \times 10^{16} \times
\left(\frac{M}{M_J}\right)^{-2}
\left(\frac{T}{7}\right)^{3}    $$

The column density is 

$$ N_{H}({\rm m}^{-2}) = 2.55 \times 10^{29} \times
\left(\frac{M}{M_J}\right)^{-1}
\left(\frac{T}{7}\right)^{2}    $$

With standard dust properties, we then expect the optical extinction to be
$A_V \sim 13,000 (M/M_J)^{-1}$, i.e. the clouds are completely opaque to normal
starlight. 

\subsection{Are the clouds optically thin or thick ? } \label{thin?}

Are the clouds optically thin (to their own submm radiation) as I have
been assuming ? We can check this by calculating the blackbody radiation from a body
at $T=7K$ assuming the virial radius derived above. The ratio of the predicted
optically thin dust luminosity to the predicted blackbody luminosity is found
to be

$$ \frac{L_{thin}}{L_{blackbody}} = 0.40 \times
\left(\frac{M}{M_J}\right)^{-1}
\left(\frac{T}{7}\right)^{2}    $$

For large clouds the optically thin assumption is justified. The smallest
clouds we have been considering ($\sim 0.1 M_J$) on the other hand are clearly
optically thick. They would better be modelled as blackbodies, but in fact as
we have seen in section \ref{temp}, a blackbody SED is marginally ruled out. For
masses around $1 M_J$, the correct SED will require radiative transfer to
calculate properly, and could for example look like a blackbody at 450\um\ but a
greybody at 1350\um . For the remainder of this paper I make the simplifying
assumption that the clouds are optically thin at $M>M_J$ and black-body like
at $M<M_J$. I now recalculate the cloud distance, and local Galactic
mass density and number density, for clouds with $M<M_J$ using the blackbody
formula, and assuming the virial size derived above.

$$ D({\rm pc}) = 149 \times \left(\frac{M}{M_J}\right)
\left(\frac{7}{5} \frac{T-2}{T}\right)
\left(\frac{F}{2 {\rm mJy}}\right)^{-1/2} \hspace{0.5in} M<M_J$$

$$
 \rho(M_\odot {\rm pc}^{-3}) = 0.0028 \times 
\left(\frac{M}{M_J}\right)^{-2}
\left(\frac{F}{2 {\rm mJy}}\right)^{3/2}
\left(\frac{5}{7}\frac{T}{T-2}\right)^{3}
\left(\frac{N}{1000}\right) \hspace{0.5in}  M<M_J                       
$$

$$ n({\rm pc}^{-3}) = 2.97 \times 
\left(\frac{M}{M_J}\right)^{-3}
\left(\frac{F}{2 {\rm mJy}}\right)^{3/2}
\left(\frac{5}{7}\frac{T}{T-2}\right)^{3}
\left(\frac{N}{1000}\right)   \hspace{0.5in}  M<M_J     $$

The density has a much steeper mass dependence than before. The main
conclusion is that if we are not to exceed the limit on local unseen mass,
$\rho < 0.03$, the mass limit becomes tighter than before : $M>0.3 M_J$. 

\subsection{Angular size and covering factor} \label{distbn}

In the previous section we found that the clouds will be opaque at visible
wavelengths. Will they produce extinction effects that should have been
previously noticed ? Taking our derived virial size and the distance deduced
for an object with a flux of 2 mJy, I find an an angular radius as follows :

$$ \theta(\arcsec) = 0.10 \times 
\left(\frac{M}{M_J}\right)^{1/2}
\left(\frac{\mu}{100}\right)^{1/2}
\left(\frac{F}{2 {\rm mJy}}\right)^{1/2}
\left(\frac{7}{T}\right)\left(\frac{5}{T-2}\right) 
\hskip 1.0cm \hspace{0.5in}  M>M_J       $$

$$ \theta(\arcsec) = 0.07 \times 
\left(\frac{F}{2 {\rm mJy}}\right)^{1/2}
\left(\frac{5}{T-2}\right) 
\hskip 1.0cm \hspace{0.5in} M<M_J       $$

The two versions are for the optically thin and optically thick limits
respectively. Next we can calculate the covering factor of all sources down to
a flux of 2 mJy, assuming a uniform distribution in space, and normalising to
surface number counts $N=1000$ deg$^{-2}$ :

$$ f = 2.5 \times 10^{-6} \times
\left(\frac{M}{M_J}\right)
\left(\frac{\mu}{100}\right)
\left(\frac{F}{2 {\rm mJy}}\right)
\left(\frac{N}{1000}\right)
\left(\frac{7}{T}\right)^2\left(\frac{T-2}{5}\right)^{-2} 
\hspace{0.5 in} M>M_J       $$

$$ f = 1.0 \times 10^{-6} \times
\left(\frac{F}{2 {\rm mJy}}\right)
\left(\frac{N}{1000}\right)
\left(\frac{T-2}{5}\right)^{-2} 
\hspace{0.5 in} M<M_J       $$

Thus large clouds ($M\sim 10^4 M_J \sim M_\odot$) would produce extinction
features 10\arcsec\ in size, covering 1\% of the sky. This would certainly have
been noticed. Small clouds ($M \sim M_J$) would be a tenth of an arcsec in size,
covering only a millionth of the sky. Historically, such features would easily
have been missed, producing occasional indentations in diffuse background
light of depth no more than a few percent. In HST imaging however, they
might appear as complete black spots, at a frequency of one or two per
WFPC2
image. Note that the angular size deduced here is only a crude estimate - as
well as the uncertain parameters quoted in the formulae above, the use of
virial size is only a first crude guess. We might expect that the angular size 
is reliable to a factor of a few. The encouraging result then is that we are
just within the testable regime. Searching for black spots may heavily
constrain or even rule out the model.

The above calculations pertain to sources down to 2 mJy. Above this flux,
about once per square degree, we would find a source at 0.2 Jy that produces
an extinction feature 1\arcsec\ across. Somewhere over the whole sky there
could be an object 20\arcsec\ across. Below 2 mJy, the background limit tells
us that the clouds could extend up to three times further at high latitude.
There might be around five per WFPC2 field, but only 0.02 - 0.03\arcsec\
across,
so that even to HST the spots would not be black. 

\subsection{Passive radio emission} \label{radio}

As discussed in section 2.3, it is possible though not certain that blank
field submm sources are very faint radio sources, with a submm/radio ratio
much larger than for low-redshift starbursts. For our template
source HDF~850.1, the marginal
detection at 8.6 GHz gives a ratio \mbox{S(850\um )/S(8.6 GHz)} = 933$\pm$
200,
and
the
upper limit at 1.4 GHz gives \mbox{S(850\um )/S(1.4 GHz)} $\geq 304$. Can
cold
dark clouds
produce weak radio emission ? The interstellar medium is of course pervaded by
cosmic rays. These produce synchrotron radiation with a volume emissivity
$\epsilon_\nu \sim 2.4 \times 10^{-41}$ W Hz$^{-1}$ m$^{-3}$ (estimated from
Fig 18.15 of Longair (1994)). Synchrotron emissivity scales as $B^{1+\alpha}$, so
perhaps enough radio emission could be produced if the magnetic field in our
clouds is significantly enhanced above the ISM average. If the heating of the
clouds is by cosmic rays, they will also maintain a steady ionisation level,
which could in turn maintain a magnetic field. We have no real way of knowing
what this field might be, but try two guesstimates. First, equipartition : if
the thermal energy density of the clouds ($n_{H_2} \times 3kT/2$) comes into
equilibrium with magnetic energy density ($B^2/2\mu_0$) then we find a predicted
field  $B_{eq}(\rm{Tesla}) = 4.0\times 10^{-6} \times (M/M_J)^{-1} (T/7)^2$.
Second, if we assume that $B\propto \sqrt{n}$ as seems to be the case for
molecular clouds, we can scale from the average density and magnetic
field in the Galactic disk, $n_H=3\times10^{6}$ m$^{-3}$ and $B=3\times
10^{-10}$ T (see Longair 1994) to find $B = 5.1 \times 10^{-5} \times
(M/M_J)^{-1} (T/7)^{3/2}$. 

Taking the smaller of these estimates, and noting that at 8.6 GHz in the ISM
$\alpha\sim 1$, we can calculate the radio luminosity to compare to the
optically thin dust luminosity for the same mass. I find the predicted
850\um\  to 8.6 GHz ratio to be :

$$ \frac{L_{850}}{L_{8.6}} = 38,156 \times 
\left(\frac{\mu}{100}\right)^{-1}
\left(\frac{T-2}{5}\right)^2
\left(\frac{T}{7}\right)^{-1} $$

Note that this result is independent of mass. The prediction fails the
goal by
a factor of 40. Using the optically thick limit for the submm emission for
small clouds improves agreement by a factor of a few. If we had used the second
field estimate ($B\propto\sqrt{n}$), we would have overshot in the other
direction by a factor of 4. On the other hand, even the equipartition field
may be too optimistic. In other words, the uncertainties are even larger than
elsewhere in this paper. Even so, there is no a priori reason why we might not
have been ten orders of magnitude out, so getting anywhere close is intriguing.
Further work may either improve or help to reject the model.

\subsection{Summary assessment of hypothesis} \label{summary}

Table 3 is a simplified summary of the various key parameters and
constraints I have derived for cloud masses ranging from $0.1M_J$ to $10^4
M_J$, assuming a temperature $T=7K$, a flux of 2mJy, a dust-to-gas ratio
$\mu=100$, and a sky density of $N=1000$ deg$^{-2}$. Over this
mass range the cloud size ranges from 1 AU to 5 pc, based on an assumption of
virial equilibrium maintained by cosmic ray heating, and the distances of such
objects range from 15 pc to 10 kpc. There are three factors however that
constrain the allowable masses :

(i) The implied local mass density rules out very small clouds, $M<0.3 M_J$,
if they are not to exceed the robust limit on local unseen matter. 
v
(ii) The absence of very obvious extinction holes at high galactic latitude
rules out very large clouds, $M>100 M_J$.

(iii) If the FIR-mm background is not to be exceeded, the clouds cannot extend
more than roughly three times further than the distance deduced for 2 mJy
sources. This mildly rules out clouds in the $M\sim 5 - 500 M_J$ range, but
is consistent with small clouds, $M\sim M_J$, being a disk population. 

All the above is for the fiducial values of the secondary parameters. Given the
uncertainties, however, we might perhaps conclude that the allowable mass
range is $M \sim 0.1 - 10 M_J$. The cold dark cloud hypothesis has not been so
easy to dismiss as brown dwarfs and comets. The various constraints have
eventually come close to ruling out the hypothesis, but an interesting mass
range remains allowed. Improved observations should be able to fairly
conclusively either dismiss or confirm the idea, as discussed further in
section \ref{test}. 

\section{Other work on cold dark clouds.}

So far I have tried to examine the local hypothesis for SCUBA sources strictly
from the viewpoint of the submm data. However over the last few years there
has been increasing observation and speculation concerning cold dark clouds.

\subsection{Small-scale structure in the ISM} \label{ISM} 

ISM structure on very small scales has been seen in several ways (see
review by Heiles 1997). The nature
of each of these structures differs significantly from the clouds we have
postulated here, but there may be a way to relate them.
AU-scale structure in {\em HI absorption} has been shown both
directly by VLBI imaging of bright background quasars (Dieter, Welch and Romney
1976;  Davis, Diamond and Goss 1996) and by the time variations of the HI
absorption spectra seen towards high-velocity pulsars (Frail \et\ 1994).
The
HI column concerned is of the order $N_H\sim 10^{24}$ m$^{-2}$, and so the total
estimated mass is of the order $10^{-8} M_\odot$, orders of magnitude smaller
than the clouds we have been discussing. Another line of evidence for AU-scale
structure is the {\em Fiedler clumps} or \lq\lq extreme scattering
events\rq\rq\  (ESEs) in quasar radio light curves (Fiedler \et\ 1987; Romani,
Blandford and Cordes 1987; Fiedler \et\ 1994). These are occasional erratic
excursions in radio flux which last a few months, and are thought to be
due to spatial variations in refractive index in intervening material.
Standard modelling of these events require ionised gas at a temperature around
10,000K, obviously rather different from our dense cold clouds. (But see
section 6.3). A key feature of both the HI and Fiedler structures is their
large
covering factor. At low Galactic latitudes at least, the HI variations in
pulsars seem to occur in essentially all cases. At high latitudes, the frequency
of ESEs indicates a covering factor somewhere in the range $10^{-3} - 10^{-5}$
(Fiedler \et\ 1994; Walker and Wardle 1998). The next piece of
observational support is the existence of small
extremely
optically thick structures in the Milky Way. Optically dark patches
are well known of course, but the new feature is compact features opaque in
the mid-IR, seen both with ISO (Perault \et 1996) and with the MSX experiment
(Egan \et 1998; Carey \et 1998). They are very cold, not being seen in IRAS
100\um\ emission. However they are not the same as the clouds
discussed in this paper. They are resolved, with $\sim$ 30\arcsec\ angular
scale, 1-5 pc physical scale, at distance 2-5 kpc, and with mass $10^5 M_\odot$.
Their sky density in the Galactic plane is around $N=20$ deg$^{-2}$, and they
are probably not seen at high latitudes, as they are not seen towards the LMC
in the MSX data (S.Price, private communication).

\subsection{Very cold dust emission in the Galaxy and elsewhere}

Reach \et\ (1995) analysed the COBE-FIRAS spectra over the whole sky and found
that they were well fitted by the sum of two greybodies - a \lq\lq
warm\rq\rq\ component with $T = 16 - 21 K$ and a \lq\lq very cold\rq\rq\
component with $T = 4 - 7$. The very cold emission is present at high latitudes
as well as in the plane. Reach \et\ argue against very cold dust clouds,
suggesting instead emission from very small grains out of equilibrium with the
ISM. Sciama (2000) argues for an origin in molecular line emission.
When deriving the isotropic extragalactic
background, Puget \et\ (1996) and
Fixsen \et\ (1998) attempt to remove the Galactic contribution of course, but
the various methods employed all assume either that the angular distribution
follows some other well known component, or that the spectral shape of the
Galactic dust emission is the same at all latitudes. Reach \et\ actually show
convincingly that the latter is not the case, so it may not not be too
surprising if some of the ubiquitous very cold component is left inside the
derived isotropic background. Krugel \et\ (1999) have argued that a  cold
dust component ($T=10 K$) is also present in the SEDs of several external
galaxies, by comparing ISOPHOT far-IR and ground-based mm measurements. The
amount of power in the cold component seen by Krugel \et\ is however much much
larger than seen by Reach \et\ in the Galactic neighbourhood. For the external
galaxies, the cold component dominates longwards of about 200\um , whereas in
the Reach \et\ fits (and in the spectral decomposition of the isotropic
background I have shown here), the cross-over point is around 1mm. Possibly the
contribution of dense cold clouds is modest in the solar neighbourhood, but
dominates the outer disks of galaxies, as argued by Pfenniger and Combes (1994).
As well as being detectable in emission, it is conceivable that such cold
dark clouds make galactic disks opaque. At the solar radius, the covering
factor of our clouds will be small (see section 5.5) but possibly it
becomes significant in the outer disks. In addition if the clouds are
actually part of a fractal ISM, as argued by Pfenniger and Combes (1994),
then the overall opacity could be much larger.

\subsection{Theoretical precedents}  \label{theory} 

Some authors have previously suggested that halo dark matter could reside in
the form of cold dark clouds and so be undetected. Pfenniger and Combes (1994)
and Pfenniger, Combes and Martinet (1994) argued that such clouds would be at or
near the traditional hierachical fragmentation limit (a few Jupiter masses), and
further argued that flat rotation curves could be explained with massive outer
disks dominated by such molecular clumps. Gerhard and Silk (1996) argued for
$1 M_\odot$ clumps spread through the halo, suggesting that otherwise too
much $\gamma$-ray background would be produced. Very recently however,
Kalberla \et\ (1999) have successfully modelled the EGRET high-latitude
emission with a halo containing Jupiter mass clumps of size 6 AU, totalling
$10^{11} M_\odot$.  De Paolis \et (1995) and Draine (1998) have argued
that halo micro-lensing
events could be due to molecular clouds.
Walker and Wardle (1998) arrived at similar characteristics by
modelling the
ESEs. One problem with these events is that the ionised
gas deduced is at a pressure orders of magnitude higher than the interstellar
medium. The orthodox solution is to assume that the structures concerned are
transient, such as knots in supernova remnants. Walker and Wardle instead
suggested that the ESEs arise in an ionised wind from a more massive object.
The several-AU size scale is set by the timescale of ESEs plus an assumed halo
velocity of 500 km$^{-1}$, and the mass of around $1 M_J$ deduced by assuming
virial equilibrium at a temperature of a few K. The covering factor of ESEs
then leads to a large total mass in such clumps, consistent with the dark halo.
Most recently Sciama (2000) has calculated the cosmic ray heating and
FIR emission of such clouds, on the assumption that they are dustless (and
so arriving at a much lower predicted luminosity than discussed in this
paper).

\section{Have we seen the dark matter ?}  \label{DM-?} 

The mass range to which our putative clouds are constrained implies a
local mass density which, while not
being dynamically dominant locally, is about that expected for the local dark
halo contribution. However, I have argued that such clouds cannot extend
further than a few hundred pc without violating the background. Furthermore, if
they were to extend through the halo, they would produce a much larger
covering
factor of extinction features. Walker and Wardle (1998) were well aware
that extinction effects are the main argument against their hypothesis, and
suggested that either the ESE clouds are dust free, or possibly that they are so
cold that the dust grains have settled into a rocky core. However the objects
postulated in this paper cannot be dust-free or we wouldn't see them. It
is tempting
to speculate that there are two populations of dense cold clouds. Population-I
is a disk population, containing dust, heated by cosmic rays, and produces the
SCUBA sources. Population-II is a halo population, and is either primaeval with
no dust, or with a rocky core, and produces the ESEs. 

Micro-lensing studies show evidence for dark bodies at
around a few tenths of a solar mass, and seem to strongly rule out objects in
the mass range we are considering here (Alcock \et\ 1996). However, this only
applies to compact objects. For Jupiter mass objects lensing much more distant
objects, the Einstein radius is about $0.03$ AU for a lens at $100$ pc, and
still only $0.3 AU$ for a lens at 10kpc. As this is much smaller than the
object size, no significant amplification will occur for cold dark clouds
within our own halo. However, the same objects in the halos of external
galaxies will appear compact and could lens background quasars (Walker 1999).
Finally, it has been suggested that {\em gas lensing} by small clouds  can
produce the kind of stellar amplification events seen in large stellar
monitoring
programmes (Draine 1998), and that conceivably objects with Walker-Wardle like
parameters (i.e. Jupiter mass and a few AU in size) can produce the entire
event rate.

\section{Cold Dark Clouds as failed stars.}  \label{f-stars} 

Regardless of their contribution to the dark matter problem, cold dark clouds
are very interesting new objects, and it will be important to confirm whether
they exist. The mass range concerned, around a Jupiter mass, is close to the
traditional fragmentation limit for collapsing clouds set by internal opacity
(Hoyle 1953; Lynden-Bell 1973; Rees 1976). Modern hydrodynamic simulations of
star formation seem to show that the fragmentation limit is set by the
turbulence scale, and is rather larger, but depends on physical conditions
(e.g. Padoan and Nordlund 1999). The objects we are discussing are not as
compact as stars or brown dwarfs, but far more compact than molecular clouds.
Viewed this way in the context of star formation they seem to be alternative
end-points for collapsing clouds. But is it reasonable that such objects
could exist ? Here I look briefly at three problems. (i) Is there a
heating source ? (ii) Are they stable ? (iii) Will they survive
disruption ?

Once a protostellar cloud becomes
opaque to heating by external starlight, it can collapse, and as it radiates
heat away it collapses further and gets hotter. This process continues until a
new source of support emerges. For massive enough clouds, this is of course
nuclear fusion in the centre producing a source of heat and pressure. For
a small cloud, we have seen in section 5.3 that the cooling time is
very short, so unless there is an external heating source, it will
continue to collapse until becoming degenerate - i.e. a brown dwarf. The
heating source that could prevent this is cosmic rays, as they will reach
deep into even these very thick clouds. The local CR energy density
is quite significant, $\epsilon_{CR} \sim 1.80$ eV cm$^{-3}$ (Webber
1998), but it is not clear what fraction of this is available for heat.
Sciama (2000) has considered the CR heating of Walker-Wardle-like clouds,
but assumed that cooling was by molecular line emission. Here I assume
cooling by
dust emission. Integrating over a greybody spectrum with $\beta=1.3$,
assuming the parameters of a one Jupiter mass cloud as derived in this
paper, and assuming that 1\% of the CR energy density is available for
heat, I arrive at an equilibrium temperature of $T=9.5$, very much in the
range we are looking for. This depends very weakly on most of
the parameters concerned, including the efficiency of heating (roughly to
the power 0.1) 

But will a CR heated equilibrium be stable ? The sound crossing time is
very short ($\sim$ 100 years) so the objects are stable against pressure
perturbations and
should find a hydrostatic equilibrium equivalent to the virial equilibrium we
have been assuming. However this equilibrium may be unstable on the cooling
timescale (10$^4$ years). If the object collapses, it will heat up and
radiate faster than the cosmic rays can re-supply the energy, leading to
further collapse. This lack of thermal stability is certainly the major
theoretical
objection to the cold dark cloud picture. Wardle and Walker (1999)  
discuss ways
to circumvent this problem by the sublimation of solid hydrogen. The possibility
of magnetic or rotational support should also be investigated. Finally, it
is possible our
clouds are not long lived objects, but part of a dynamic interstellar medium in
which clouds are repeatedly destroyed and re-formed (Pfenniger and
Combes 1994).

Will the clouds be disrupted by passing stars ?   The tidal disruption
radius for a star of mass $M_*$ will be
given by 

$$ D_{\rm disrupt} = 316 {\rm AU} \times
\left(\frac{M_*}{M_\odot}\right)^{1/2} 
\left(\frac{M_c}{M_J}\right)^{1/2}
\left(\frac{T}{7}\right)^{-1} $$

Taking a mean stellar mass of $0.5 M_\odot$, a density of $n=0.1$
pc$^{-3}$, and a typical random velocity  of $v=30$ km
s$^{-1}$, the two-body collision timescale gives 

$$ \tau_{\rm disrupt} = 8.4 \times 10^{10}{\rm years} \times
\left(\frac{M_c}{M_J}\right)^{-1}
\left(\frac{M_*}{0.5 M_\odot}\right)^{-1} 
\left(\frac{T}{7}\right)^{2} $$

The conclusion is that only low mass objects
are long lived. Objects more massive than $100 M_J$ have lifetimes less than the
age of the Galactic disk. Re-assuringly, the massive objects which we earlier
concluded cannot dominate the SCUBA counts without violating observational
constraints, are also those which we do not expect to survive in large numbers.
We can likewise calculate a cloud-cloud collision timescale, which gives a
similar timescale, 
$\tau = 9.7 \times 10^{10}{\rm years}\times \left(M/M_J\right)^{-1/2}$.
Gerhard and Silk (1996) calculate that a cloud can only survive such a
collision if $N_H>10^{29}{\rm m}^{-2}$, again suggesting that only low
mass clouds can survive. Finally we might ask how often a typical star
(like the Sun) should wander through such a completely opaque cloud - is
this what killed the dinosaurs ??
Given the cloud size and population density for our clouds derived in
earlier sections, I find that such encounters are very rare, 
$\tau = 3.9 \times 10^{11}{\rm years}\times \left(M/M_J\right)^{-1/2}$,
although the probability of such an encounter may be significantly
enhanced when the sun crosses a spiral arm (see Leitch and Vasisht 1997).

However stars actually form, stars are stable against disruption and stellar
mass cold clouds are not. Planetary mass cold clouds are stable against
disruption, but it remains to be seen whether they are thermally stable. The 
contentious issue then is whether nature forms cold dark clouds or brown
dwarfs, and how it chooses.

\section{Testing the hypothesis}  \label{test} 

Several lines of investigation look promising. 

(i) Looking for unresolved or marginally resolved dark spots against diffuse
background light sources. So far we have restricted the allowed mass range on
the basis that no such gross effects are known, but a more careful search
against carefully chosen sources is obviously feasible. Furthermore any such
dark spots should coincide with bright spots in submm maps. 

(ii) Looking for stellar switch-on-offs. Given the deduced covering factor, about
one background star in every million should be occulted at any one time. With a
size of $\sim 10$ AU and a random disk velocity with respect to the sun of say
$30$ km s$^{-1}$, the occultation would last a year or two. Such rare
but dramatic appearances and disappearances may just be detectable with
existing datasets such as those of the MACHO or OGLE projects.

(iii) Over a period of years the submm sources should show measurable proper
motion. This won't be detectable with SCUBA, but should be with IRAM. If the
cold dark clouds are also radio sources, then the VLA sources should also show
proper motion, and given the accuracy of radio positions this may be an easier
project.

(iv) Zero redshift molecular emission lines may be measurable. At a temperature
of $T=4-9 K$ $H_2$ will not be excited, but the lowest states of CO are at
$2-3K$, so should be excited. I have not
attempted to calculate the expected line fluxes, given the uncertainty in
conditions and optical depth. Downes \et\ (1999) made a sensitive line search
in HDF~850.1 but only in narrow windows where {\em redshifted} lines would be
expected.

(v) As suggested by Walker and Wardle (1999), clouds relatively near stars may
show up as H$\alpha$ sources from the ionised winds. The scattered starlight
might also be detectable.

(vi) Somewhere on the sky a dark object with a submm flux of many Jy and
and a size of 30\arcsec\ may be lurking, somewhere on the edge of the Oort
Cloud. Possibly such an object has already been detected by its effect on
cometary orbits (Murray 1999).

(vii) Source counts should show a moderate dependence on Galactic latitude,
depending on the scale height of the population.

(viii) The fraction of clearly identified sources should depend on wavelength.
At 450 \um\ most sources would be identfied with galaxies at redshift $\sim 1-3$.
At 1.3 mm most sources would be blank.

(ix) At bright counts (brighter than about 10 mJy) source counts should be
Euclidean.

\section{Conclusions}  \label{conc} 

The lack of clear optical identifications for a large fraction of faint submm
sources is pushing us towards believing that they are very high redshift
objects, but we must first scrupulously consider alternative possibilities. In
local hypotheses, these sources are extremely cold, $T\sim 7$ K. They cannot be
brown dwarfs or solar system bodies without grossly violating simple
constraints. Cold dusty clouds are harder to rule out completely, but if such
objects are not to violate limits on local mass density, extinction features,
and the FIR background, they are constrained to have masses similar to Jupiter
to within an order of magnitude, and to be a Galactic Plane population. Such
clouds do not explain the dark matter problem, unless there is a much larger
population of similar but dust-free objects. (In this respect my proposal
differs from most of the rest of the literature in this area which
involves a pervasive but dust-free halo population). If such cold dusty
clouds do
exist, they are an important new component of the interstellar medium,
and may be an important part of the star formation puzzle. The main theoretical
objection is that such objects should be unstable to collapse on a thermal
timescale, but they may be short-lived objects which are part of a dynamic
ISM, as argued by Pfenniger and Combes (1994).

The majority of faint SCUBA sources clearly really are distant galaxies.
However if even a quarter or a third of the SCUBA sources are actually local
objects, the cosmological implications are significant, as this would
selectively remove the objects believed to be at $z>3$.

\section{Acknowledgements}  \label{acknow} 

The idea to examine local hypotheses for SCUBA sources came from a
sceptical lunchtime remark by Adrian Webster. (He claims to have forgotten
this !) Since then, I have had many very useful conversations
with a large number of people. In particular Nigel Hambly taught me about
brown dwarfs, told me about the work of Walker and Wardle, and generally
encouraged and helped me. In addition, key points that have ended
up in the paper were provoked by Peter Brand, Raul Jimenez,
Donald Lynden-Bell, Richard Hills, Jacco Van Loon and Mark Walker. Very
useful critical readings by Michael Rowan-Robinson and Rob Ivison also
helped a lot.

\newpage

{\Large\bf REFERENCES}

Alcock, C., Allsman, R.A., Axelrod, T.S., Bennett, D.P., Cook, H.K., Freeman,
K.C., Griest, K., Guern, J.A., Lehner, M.J., Marshall, S.L., Park, H.-S.,
Perlmutter, S., Peterson, B.A., Pratt, M.R., Quinn, P.J., Rodgers, A.W.,
Stubbs, C.W., and Sutherland, W., 1996. \apj\vol{461}{84}

Almaini, O., Lawrence, A., and Boyle, B.J., 1999. \mn\vol{305}{L59}

Barger, A.J., Cowie, L.L., Sanders, D.B., Fulton, E., and Taniguchi, Y.,
Sato, Y., Kaware, K., and Okuda, H, 1998. \nat\vol{394}{248} 

Barger, A.J., Cowie, L.L., and Sanders, D.B., 1999. \apj\vol{518}{L5}

Binney, J., and Merrifield, M., 1998. {\it Galactic Astronomy}, chapter 9.
Published by Princeton University Press.

Blain, A.W., 1999. \mn\vol{309}{955}

Blain, A.W., Kneib, J.-P., Ivison, R.J., Smail, I., 1999a. \apj\vol{512}{L87}

Blain, A.W., Smail, I., Ivison, R.J., and Kneib, J.-P., 1999b. \mn\vol{302}{632}

Burrows, A., Marley, M., Hubbard, W.B., Lunine, J.I., Guillot, T., Saumon, D.,
Freedman, R., Sudarsky, D., and Sharp, C., 1997. \apj\vol{491}{856}

Carey, S.J., Clark, Egan, M.P., Price, S.D., Shipman, R.F., and Kuchar, T.A.,
1998. \apj\vol{508}{721}

Carilli, C.L., and Yun, M.S., 1999. \apj\vol{513}{L13}

Creze, M., Chereul, E., Bienayme, O., and Pichon, C., 1998. \AnA\vol{329}{920}.

Davis, R.J., Diamon, P.J., and Goss, W.M., 1996. \mn\vol{283}{1105}

De Paolis, E., Ingrosso, G., Jetzer, Ph., and Roncadelli, M., 1995.
\AnA\vol{295}{767}

Dieter, N.M., Welch, W.J., and Romney, J.D., 1976. \apjl\vol{206}{L113}

Downes, D., Neri, R., Greve, A., Guilloteau, A., Casoli, F., Hughes, D., Lutz,
D., Menten, K.M., Wilner, D.J., Andreani, P., Bertoldi, F., Carilli, C.L.,
Dunlop, J., Genzel, R., Geuth, F., Ivison, R.J., Mann, R.G., Mellier, Y.,
Oliver, S., Peacock, J., Rigopoulou, D., Rowan-Robinson, M., Schilke, P.,
Serjeant, S., Tacconi, L.J., and Wright, M., 1999. \AnA\vol{347}{809}.

Draine, B.T., 1998. \apjl\vol{509}{L41}

Dunne, L., Eales, S., Edmunds, M., Ivison, R., Alexander, P., and
Clements, D.L., 2000. Astro-ph/0002234.

Dwek, E., Arendt, R.G., Hauser, M.G., Fixsen, D., Kelsall, T., Leisawitz, D.,
Pei, Y.C., Wright, E.L., Mather, J.C., Moseley, S.H., Odegard, N., Shafer, R.,
Silverberg, R.F., and Weiland, J.L., 1998. \apj\vol{508}{106}

Eales, S., Lilly, S., Gear, W., Dunne, L., Bond, J.R., Hammer, F., Le fevre,
O., and Crampton, D., 1998. \apj\vol{515}{000}

Eales, S., Lilly, S., Webb, T., Dunne, L., Gear, W., Clements, D., and
Yun, M., 2000. \aj in press. Astro-ph/0009154 

Egan, M.P., Shipman, R.F., Price, S.D., Carey, S.J., Clark, F.O., and Cohen,
M., 1998. \apjl\vol{494}{L199}

Evans, N.W., 1996. \mn\vol{278}{L5}.

Fabian, A.C., Smail, I., Iwasaw, K., Allen, S.W., Blain, A.W., Crawford,
C.S., Ettori, S., Ivison, R.J., Johnstone, R.M., Kneib, J.-P., and WIlman,
R.J., 2000. \mn\ in press.

Fiedler, R.L., Dennison, B., Johnston, K.J., and Hewish, A., 1987.
\nat\vol{326}{675}

Fiedler, R., Dennison, B., Johnston, K.J., Waltman, E.B., and Simon, R.S.,
1994. \apj\vol{430}{581}

Fixsen, D.J., Dwek, E., Mather, J.C., Bennett, C.L., and Shafer, R.A., 1998.
\apj\vol{508}{123}

Frail, D.A., Weisberg, J.M., Cordes, J.M., and Mathers, C., 1994.
\apj\vol{436}{144}

Frayer, D.T., Ivison, R.J., Scoville, N.Z., Yun, M., Evans, A.S., Smail,
I., Blain, A.W., and Kneib, J.-P., 1998. \apj\vol{506}{L7}.

Frayer, D.T., Ivison, R.J., Scoville, N.Z., Evans, A.S., Yun, M., 
A.S., Smail, I., Barger, A.J., Blain, A.W., and Kneib, J.-P., 1999.
\apj\vol{514}{L13}.

Gerhard, O., and Silk, J., 1996. \apj\vol{472}{34}

Heiles, C., 1997. \apj\vol{481}{193}

Holland, W.S., Robson, E.I., Gear W.K., Cunningham, C.R., Lightfoot, J.F.,
Jennes, T., Ivison, R.J., Stevens, J.A., Ade, P.A.R., Griffin, M.J., Duncan,
W.D., Murphy, J.A., and Naylor, D.A.,  1999. \mn\vol{303}{659} 

Holmberg, J., and Flynn, C., 2000. \mn\vol{313}{209}

Hoyle, F., 1953. \apj\vol{118}{513}

Hughes, D., Dunlop, J.S., and Rawlings, S., 1997. \mn\vol{289}{766}.

Hughes, D.H., Gear, W.K., and Robson, E.I., 1994. \mn\vol{270}{641}.

Hughes, D., Serjeant, S., Dunlop, J.S., Rowan-Robinson, M., Blain, A., Mann,
R.G., Ivison, R., Peacock, J., Efstathiou, A., Gear, W., Oliver, S., Lawrence,
A., Longair, M., Goldschmidt, P., and Jenness, T., 1998. \nat\vol{394}{241}.

Ivison, R.J., Smail, I., le Borgne, J.-F., Blain, A.W., Kneib, J.-P.,
Bezecourt, J., and Davies, J.K., 1998. \mn\vol{298}{583}.

Kalberla, P.M.W., Shchekinov, Yu.A., and Dettmar, R.-J., 1999. \AnA\vol{350}{L9}

Krugel, E, Siebenmorgen, R., Zota, R., and Chini, R, 1998. \AnA\vol{331}{L9}

Leitch, E.M., and Vasisht, G., 1997. \it New Astronomy, \rm\bf 3, \rm 51.

Lilly, S.J., Eales, S.A., Gear, W.K.P., Hammer, F., Le Fevre, O., Crampton, D.,
Bond, J.R., and Dunne, L., 1999. \apj\vol{518}{641}

Longair. M.J., 1994. {\it High Energy Astrophysics, Vol 2}. Published by
Cambridge University Press.

Lynden-Bell, D., 1973. In {\it Dynamical Structure and Evolution of stellar
systems}, Saas-Fee Advanced Course, eds. Contopoulos, G., Henon, M., and
Lyden-Bell, D., published by Geneva Observatory.

McMahon, R.G., Priddey, R.S., Omont, A., Snellen, I., Withington, S., 1999.
\mn\vol{309}{L1}

Murray, J.B., 1999. \mn\vol{309}{31}

Padoan, P., and Nordlund, A., 1999. \apj , in press.

Perault, M., Omont, A., Simon, G., Seguin, P., Ojha, D., Blommaert, J., Felli,
M., Guglielmo, F., Habing, H., Price, S., Robin, A., de Batz, B., Cesarsky, C.,
Elbaz, D., Epchtein, N., Fouque, P., Guest, S., Levine, D., Pollock, A.,
Prusti, T., Sibenmorgen, R., Testi, L., and Tiphene, D., 1996.
\AnA\vol{315}{L165}

Pfenniger, D., and Combes, F., 1994. \AnA\vol{285}{95}

Pfenniger, D., Combes, F., and Martinet, L., 1994. \AnA\vol{285}{79}.

Puget, J.-L., Abergel, A., Bernard, J.-P., Boulanger, F., Burton, W.B., Desert,
F.-X., and Hartmann, D., 1996. \AnA\vol{308}{L5}

Reach, W.T., Dwek, E., Fixsen, D.J., Hewagama, J., Mather, J.C., Shafer, R.A.,
Banday, A.J., Bennet, C.L., Cheng, E.S., Eplee Jr, R.E., Leisawitz, D., Lubin,
P.M., Read, S.M., Rosen, L.P., Shuman, F.G.D., Smoot, G.F., Sodroski, T.J.,
and Wright, E.L., 1995. \apj\vol{451}{188}

Rees, M.J., 1976. \mn\vol{176}{483}

Richards, E.A., 1999. \apj\vol{513}{L9}

Richards, E.A., Kellerman, K.I., Fomalont, E.B., Windhorst, R.A., Partridge,
R.B., 1998. \aj\vol{116}{1039}.

Romani, R.W., Blandford, R.D., and Cordes, J.M., 1987. \nat\vol{328}{324}

Sciama, D.W., 2000. \mn\vol{312}{33}

Smail, I., Ivison, R.J., and Blain, A.W., 1997. \apj\vol{490}{L5}

Smail, I., Ivison, R.J., Kneib, J.-P., Cowie, L.L., Blain, A.W., Barger, A.J.,
Owen, F.N., and Morrison, G., 1999. \mn\vol{308}{1061}

Smail, I., Ivison, R.J., Owen, F.N., Blain, A.W., and Kneib, J.P., 2000.
\apj\vol{528}{612}

Walker, M., and Wardle, M., 1998. \apj\vol{498}{L125}

Walker, M.A., 1999. \mn\vol{306}{504}

Wardle, M., and Walker, M., 1999. \apj\vol{527}{L109}

Webber, W.R., 1998. \apj\vol{506}{329}

Weissman, P.R., 1995. \anrev\vol{33}{327}

\newpage
\begin{table}
\begin{tabular}{lcr}
\multicolumn{3}{c}{TABLE 1 : MEASURED FLUXES OF HDF 850.1} \\
\\
Wavelength  & Flux                         & Reference \\ \hline
  15\um     & $<$23\uJy (3$\sigma$)      & (1)  \\
 450\um     & $<$21 mJy (3$\sigma$)      & (1)  \\
 850\um     & 7.0$\pm$0.4 mJy            & (1) \\ 
1350\um     & 2.1$\pm$0.5  mJy           & (1) \\
1270\um     & 2.2$\pm$0.3  mJy           & (2) \\
 2.8mm      & $<$0.5 mJy (3$\sigma$)     & (2) \\
 3.4mm      & $<0.4$ mJy (3$\sigma$)     & (2) \\
 3.5cm      & 7.5$\pm$2.2\uJy            &  (3,2) \\
 20cm       & $<$23\uJy (3$\sigma$)      &  (4,2) \\ \hline
\end{tabular}
\caption{Measured Fluxes of HDF~850.1 at various wavelengths, compiled from the
literature.  References : (1) Hughes \et\ 1998 (2) Downes \et\ (1999)
(3) Richards \et\ (1998)
(4) Richards (1999)  }
\end{table}

\vskip 1cm

\begin{table}
\begin{tabular}{lllllll}
\multicolumn{7}{c}{TABLE 2 : TEMPERATURE LIMITS FROM HDF 850.1 FLUX RATIOS} \\
\\           
ratio  &  & & $\beta=$0  & $\beta=$1 &$\beta=$1.5  &$\beta=$2 \\   \hline

S(450\um )/S(850\um )  &$<$ & 2.1           &T$<$18  &T$<$9.4  &T$<$7.6
&T$<$6.4 \\  
S(850\um )/S(1300\um ) &$=$ & 3.2$\pm$0.5   &T$<$70  &T$>$8.3
&T$=$6.0-49.6
&T$=$4.7-12.4  \\ 
S(1300\um /2800\um )   &$<$ & 5.7           & excluded  &T$>$6.5  &T$>$4.1
&T$>$3.1 \\
\\
combined   & &                      & excluded  &T$=$8.3-9.4  &T$=$6.0-7.6   
& T$=$4.7-6.4 \\ \hline

\end{tabular}
\caption{Limits on single temperature models for HDF~850.1, based on
individual flux ratios, assuming greybodies with various values of the
emissivity index $\beta$. The limits and ranges used are approximately 95\%
confidence. Note that the SCUBA 1350\um\ and IRAM 1270\um\ fluxes have
been averaged and taken as a 1300\um\ flux.} 

\end{table}

\vskip 1cm

\begin{table}
\begin{tabular}{lccccc}
\multicolumn{6}{c}{TABLE 3 : SUMMARY OF CONSTRAINTS} \\ 
\\

& 0.1 M$_J$  & M$_J$   & 10M$_J$     & 100$M_J$    &10M$\odot$ \\ \cline{2-6}

optically thick or thin ?    
& thick          & intermediate & thin    & thin  &  thin \\ 

Virial Radius    
&1 AU         &10 AU         & 100 AU  & 1000 AU   & 5pc \\ 

Extinction through cloud (A$_V$)         
& A$_V$=10$^5$   &A$_V$=10$^4$    &A$_V$=1300   & A$_V$=130  &A$_V$=1.3  \\

Population mass density (M$_\odot$ pc$^{-3}$)        
& (0.3)    & 0.01         & 0.003   & 0.001  & 10$^{-4}$ \\ 

Approx. distance limit
& 45pc        & 300pc   & (1 kpc)  & (3 kpc)  & 30 kpc  \\

Angular radius
& 0.07\arcsec & 0.1\arcsec  &0.3\arcsec & (1\arcsec) & (10\arcsec) \\

Covering factor
&10$^{-6}$     & 10$^{-6}$  & 10$^{-5}$  & (10$^{-4}$) & (10$^{-2}$) \\

\end{tabular}
\caption{Summary of deduced cloud properties as a function of assumed cloud
mass. The figures shown are rounded values, and assume (where appropriate) dust
temperature T$=7K$, dust-to-gas ratio $\mu=100$, surface number counts N=1000
deg$^{-2}$, and submm flux S(850\um = 2 mJy). The distance limit is very
approximate and comes from the requirement not to exceed half the submm
background flux. Boxes taken to be excluded by observations are bracketed} 

\end{table}

\vskip 1cm


\newpage 
%
\begin{figure}[h]
\centerline{\epsfig{file=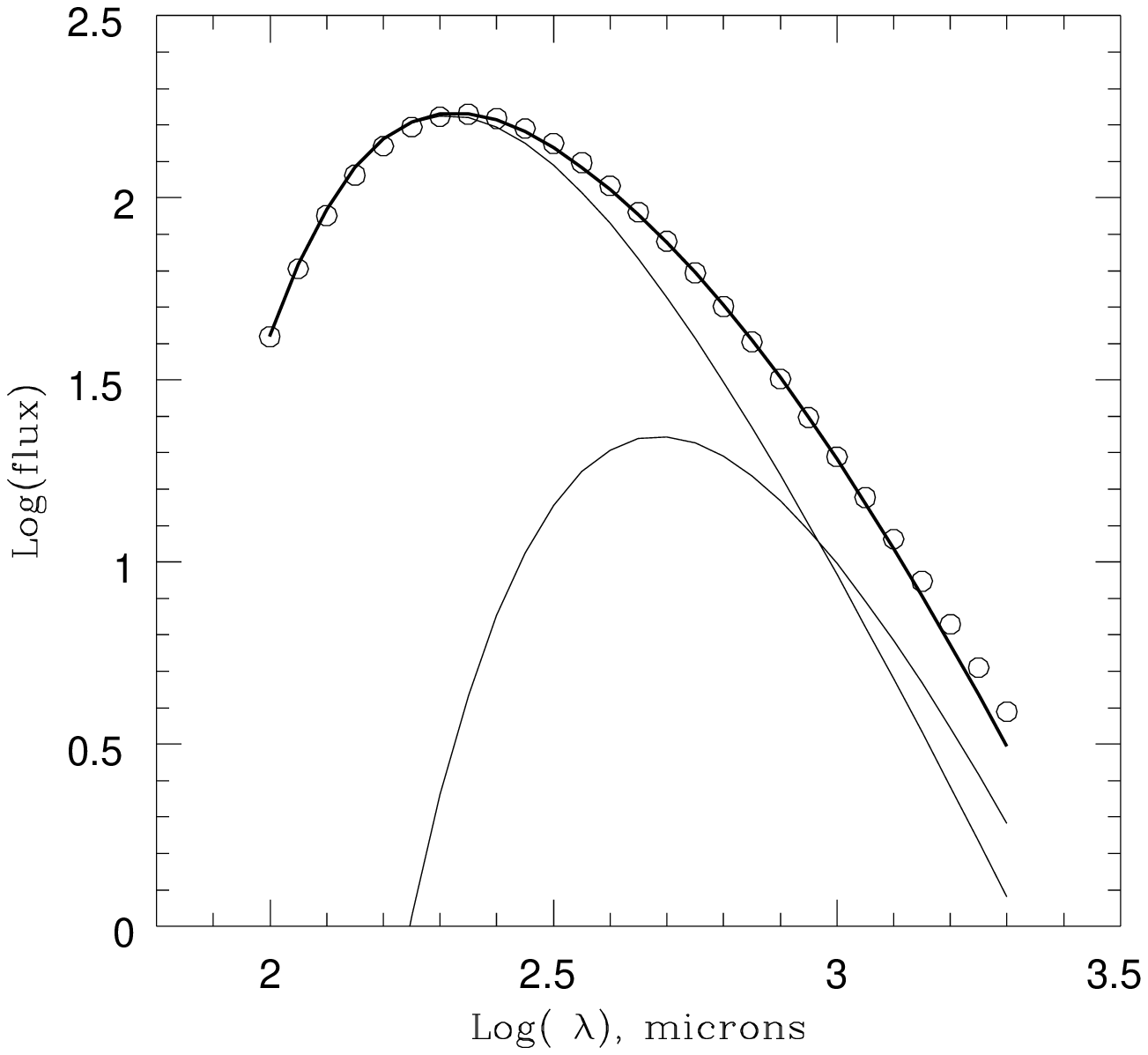,width=17cm}}
\caption[]{\small\it  \rm Decomposition of the FIR-mm
background spectrum. The actual data are not shown here. The circles represent
the parameterisation found by Fixsen \et\ to fit the data. The thin lines are
both single temperature greybody spectrum. Each has $\beta=1.3$. The upper
line has $T=17$ and the lower line has $T=7$. The thick line is the sum of the
two greybody spectra.  }  

\end{figure}

%

\end{document}